\documentclass[12pt]{article}

\usepackage{sbc-template}
\usepackage{graphicx,url}
\usepackage[english,brazil]{babel}   
\usepackage[utf8]{inputenc}  
\usepackage{color}
\usepackage[table,xcdraw]{xcolor}

\definecolor{codegreen}{rgb}{0,0.6,0}
\definecolor{codegray}{rgb}{0.5,0.5,0.5}
\definecolor{codepurple}{rgb}{0.58,0,0.82}
\definecolor{backcolour}{rgb}{0.95,0.95,0.92}

\title{In-Person and Remote Participation Review at IETF: Collaborating Without Borders}

\author{Lucas Andrade\inst{1}, Juliao Braga\inst{2,3}, Stefany Pereira\inst{1}, Rafael Roque\inst{4},\\ Marcelo Santos\inst{1} }

\address{Instituto Federal do Sertão Pernambucano (IF Sertão-PE)\\
  Salgueiro – PE – Brasil
 \nextinstitute
  Universidade Presbiteriana Mackenzie\\São Paulo, SP -- Brasil
  \nextinstitute
  Instituto Superior Técnico, INESC-ID, Universidade Lisboa, PT
\nextinstitute
  Centro de Informática (CIn)\\Universidade Federal de Pernambuco (UFPE)\\
  Recife – PE – Brasil
  \email{lucasmatheusif@outlook.com.br, stefanyp3108@gmail.com}
  \email{rrs4@cin.ufpe.br, marcelo.santos@ifsertao-pe.edu.br}
  \email{juliao.braga@tecnico.ulisboa.pt}
}

\begin{document} 

\maketitle

\begin{abstract}
  The IETF has been acting as one of the main actors when discussing standardization of protocols and good practices on the Internet. Collaborating with the IETF community can be complex and distant for many researchers and industry members because of the financial aspect to travel to the meeting. However, it notes the collaboration between industry and academia is actively and progressively developing and refining standards within the IETF. One of the incentives for the increased participation in IETF meetings is because it is being transmitted in real time since 2015, allowing for voice and chat interaction of remote participants. Thus, in this paper, we have as objectives to give a brief vision about how to collaborate with the IETF and to analyze the importance of this new form of participation of the face-to-face meetings that has been growing in recent years.
\end{abstract}
     
\begin{resumo} 
  O IETF vem atuando como um dos principais atores quando se discute padronização de protocolos e boas práticas na Internet. Colaborar com a comunidade do IETF, pode ser complexo e distante para muitos pesquisadores e membros da indústria devido ao aspecto financeiro para viajar até o encontro. No entanto, observa a colaboração entre indústria e academia se encontra de forma ativa e ascedente no desenvolvimento e aperfeiçoamento de padrões dentro do IETF. Um dos motivos pelo aumento de participação nas reuniões do IETF, é o fato de haver a transmissão em tempo real desde 2015, permitindo a interação por voz e chat de participantes remotos. Assim, neste artigo temos como objetivos dar uma breve visão sobre como colaborar com a IETF e analisar a importância dessa nova forma de participação das reuniões presenciais que vem crescendo nos últimos anos. 
\end{resumo}

\section{Introduction}

The Internet has grown frighteningly since the first email sent in 1969 at the University of California at Los Angeles (UCLA) to a computer at Stanford Research Institute. The number of home users grown up from zero at that time to more than 4 billion in 2018. In the same period, one of the first protocols called \textit{Request for Comments} (RFC) came about through the work of Vint Cerf, Steve Crocker, and Jon Postel.

The Internet is formed by a complex structure that works through the collaboration and cooperation of diverse stakeholders. Given the large number of entities and people that make up the Internet, this structure is not always clear, even for professionals working in the area of computer networks. In this sense, it is fundamental to define, structure layers of networks, as well as create and improve protocols associated with each layer to have a better functioning of the Internet. In this context, we should highlight the \textit{Internet Engineering Task Force} (IETF) group, which is a broad and open international community composed of technicians, agencies, manufacturers, suppliers and researchers, concerned with the evolution of the architecture of the Internet. Internet, through the creation of protocols in a collaborative way. Among them, we can mention several protocols such as HTTP \cite{RFC2616}, TCP \cite{RFC0793}, and IP \cite{RFC0791}.

One of the big issues behind the IETF work is how the whole evolutionary process of the Internet architecture works and how to collaborate with that community. For many researchers and industry members contribute to the IETF is something distant and difficult. However, the Internet Society\footnote{\url{https://www.internetsociety.org/}} and the IETF itself have been working to bring new members into their community. One of the incentive practices is the online transmissions of all your live discussions and the possibility of interaction via chat or voice to discuss and clarify doubts. This practice has been adopted since November 2015 (IETF 94 in Yokohama, Japan).

The contributions of this paper are summarized as follows: (1) Collect and analyze data on the remote and face-to-face participation of IETF meetings in recent years; (2) Clarify technical aspects about the operation of the IETF and how to contribute and (3) Determine the importance of the inclusion of remote participants in each IETF meeting.

The remainder of this article is organized as follows. Section 2 gives a brief overview of how the IETF / IRTF works. Section 3 discusses the main concepts about Drafts and RFCs. Section 4 describes the methodology. Section 5 displays the results and discussions. Section 6 addresses the final considerations of the work.

\section{What is IETF / IRTF: An initial view} 

The IETF is an international non-profit, self-organized community divided into sub-groups of different areas that help identify short-term Internet related issues in order to solve them, but it is also responsible for creating standards that we use every day. The \textit{Internet Research Task Force}\footnote{\url{https://irtf.org}} (IRTF), as well as the IETF, is also responsible for identifying problems on the internet but differs from worrying about long-term issues. Members of this group do research on problems that require more time and more dedicated studies. Thus, IRTF's main function is to obtain a future vision of the Internet.

The members of the IETF and IRTF, are composed of several areas such as providers and equipment manufacturers, researchers, teachers, students and others who are interested in contribute to the development of the Internet, ie, anyone with a voluntary interest in the development of the Internet. The official language, to express itself (oral and written) is the English language, which in this case is the adopted language as the communication standard among the members. The main means of communication are the mailing lists. Both groups work on creating open standards where all of them are available to the general public, royalty free so anyone can adopt the developed standards. The creation of standards is based on an approximate consensus of its members, that is, according to the consensus of the majority of members, in a democratic way. It should be noted that IETF members work voluntarily and do not impose any mandatory standards, their standards are adopted freely and often, without major oversight. In addition to the mailing lists, IETF participants meet three times a year to complement what was discussed by e-mail among their participants. The meetings take place in different countries and last for a week, and are paid for demanding a large infrastructure to serve more than a thousand people. The cost per participant, with enrollment only, is around USD 700.00, which together with the cost of tickets and lodging can prevent many members participation due to high investment.

For more details, \textbf{The Tao of the IETF}\footnote{\url{https://www.ietf.org/tao.html}} provides a broader view of how this community works \cite{Braga:2014}.

\section{Drafts and RFCs: Basic Concepts and their Structure}

A draft is an initial document for creating an RFC. This document is intended for a working group so that the community can comment on it and make improvements. Drafts are subject to any kind of changes so they can not be cited anywhere as formal documents because they are subject to removals and changes, not having positions within the IETF until they are adopted by some group, and if so, become an RFC.

Known as Internet standards, RFCs are a set of technical documents that the IETF has, detailing the protocols being proposed, but actually not all RFCs are a standard. There are six types of RFCs: \textit{Proposed Standards}, \textit{Internet Standards}, \textit{Best Current Practices}, \textit{Informational Documents}, \textit{Experimental Documents}, and \textit{Historical Documents}. When an RFC needs some change, a new RFC is generated, without it being necessary to delete the original RFC, conserving it so that the old RFC can be studied by anyone, also serving as a model for new updates.

\subsection{Writing and Submission Processes}

The IETF is organized into seven major areas. Within these areas, there are dozens of \textit{work groups} (WGs) working in parallel on various issues that guide the Internet. So, before writing any document in the IETF it is necessary to identify if it fits in with some WG so that there is an effective collaboration of the community. Before a document becomes an RFC initially it is defined only as an I-D (Internet Draft), which is published and exposed to receive comments with an expiration date of the document. The term of validity of the document may be extended as many times as the community deems necessary until the document becomes an RFC or is filed.

For a draft to change to an RFC, an \textit{Area Director} (AD) must apply for referral to the \textit{Internet Engineering Steering Group} (IESG). The AD does its own analysis and can ask for some draft adjustments before sending it to the IESG. The IESG evaluation teams are formed by the \textit{Security Area Directorate} (secdir) and \textit{General Area Review Team} (Gen-Art), which validate the draft so that it becomes an RFC. After these processes, the RFC Editor will publish the draft as an RFC.

\section{Methodology}

The face-to-face meetings of the IETF are held three times a year. The last meeting held during the writing of this article was the 100th meeting in Singapore. The IETF provides information on the participation in their website\footnote{\url{https://www.ietf.org/registration/ietf100/attendance.py}} where you can change the meeting number in the URL itself and check the list of all participants in a specific meeting. In addition, there is a database with all drafts and RFCs of the IETF and IRTF available in their website\footnote{\url{https://datatracker.ietf.org/}}, also.

After conducting a survey of the operation of the IETF website, a Web Crawler was developed to collect all IETF RFCs and drafts. At this point it is important to highlight that the authors, region, company, work group and other information of each document were identified in an automated way. Next, the data of remote and face-to-face participants of each meeting were collected.

\begin{figure}[ht]
\centering
\includegraphics[scale=0.5]{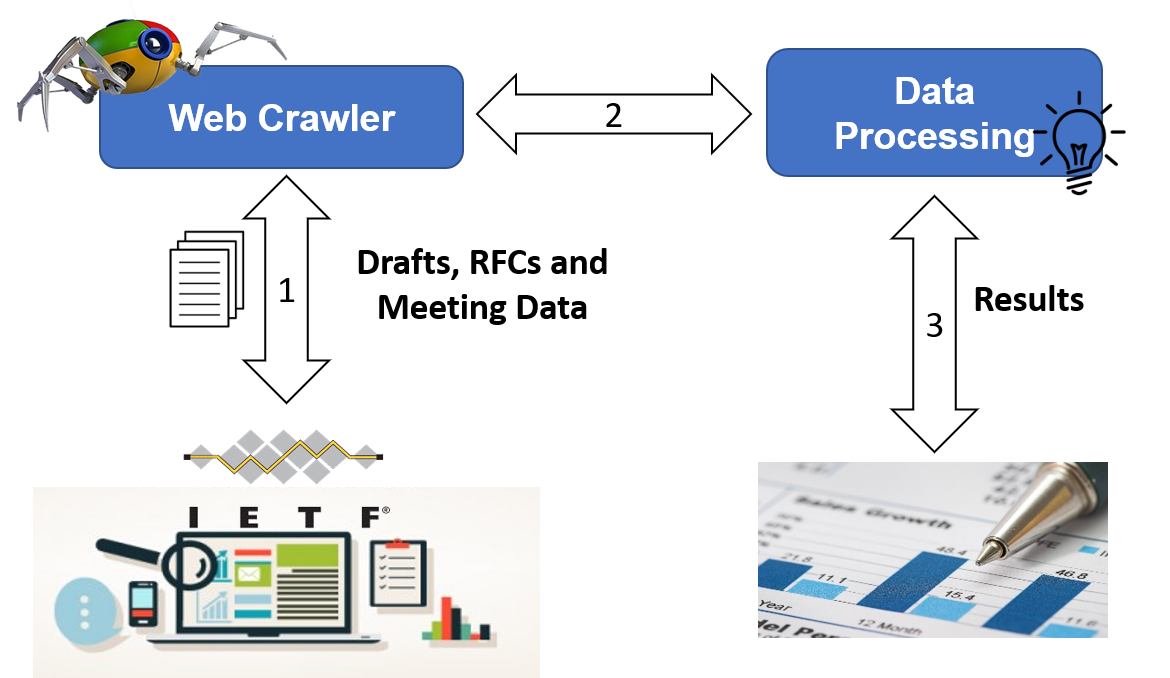}
\caption{Data Collection and Processing}
\label{figures:fig1}
\end{figure}

For the development of the Web Crawler, we used the language Java 8 and library version 1.11.2. As can be seen in Figure \ref{figures:fig1}, we initially collected all the data in step 1, then processed the data and normalized the collected content (step 2). Note that step 2, not always the name that appears in the documents (drafts and RFCs) are in the same way as the registration names in face-to-face meetings, which makes it difficult to cross information. In addition, when we do not have document data in XML format, but only text, it becomes more complex to accurately collect multiple authors in one document. In step 3, data analysis and results generation were performed.

\section{Results and Discussion}

\subsection{In-Person and Remote Participation at IETF Meetings}

\begin{figure}[ht]
\centering
\includegraphics[scale=1]{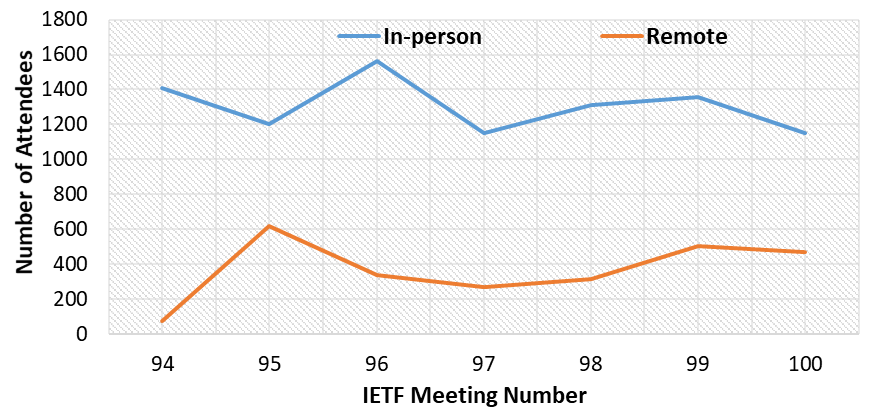}
\caption{Number of Participants per Meeting}
\label{figures:fig2}
\end{figure}

From the meeting 94  in November 2015, the IETF started broadcasting all the meetings remotely for free. In addition to having its free registration, the remote participant was able to participate and interact by voice and text. In Figure \ref{figures:fig2} we can see the number of in-person and remote attendees at each meeting. In these last eight meetings, there is no clear correlation between the number of remote and face-to-face participants. Except for the 95th meeting in Argentina, we can note that the record number of remote participants may have impacted the number of face-to-face participants.

\subsection{Participation by Countries and Continents}

\begin{figure}[ht]
\centering
\includegraphics[scale=1.0]{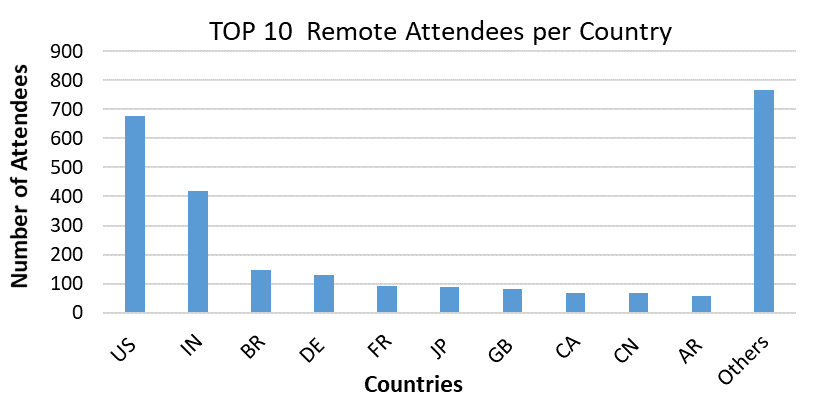}
\caption{Remote Participants by Meeting}
\label{figures:fig3}
\end{figure}

According to RFC 5646 of good practice, the IETF adopts the ISO 3166 standard to represent country code \cite{RFC5646}. Following this pattern, Figure \ref{figures:fig3} shows the 10 countries with the largest number of remote participants based on all remote meetings already held (sum of remote attendees from meeting 94 to 100).

As with the number of face-to-face participants, the largest number of remote participants is from the United States. However, countries like India, Brazil and Argentina appear respectively in the 2nd, 3rd and 10th places. It should be noted that if we added the number of participants from other countries that have few remote participants outside the top 10, the number of participants exceeds the number of participants in the United States. This shows that even in small numbers, remote participation has provided a greater plurality of participants and this is positive in a community that depends on collaboration and adoption of standards in a democratic way.

\begin{figure}[ht]
\centering
\includegraphics[scale=1.0]{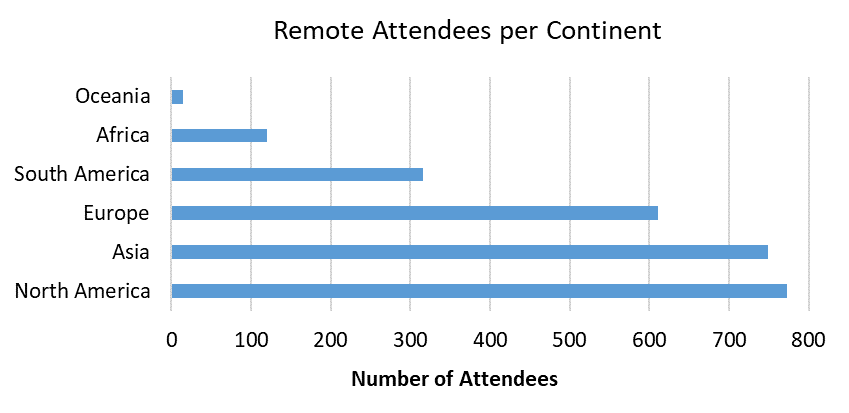}
\caption{Number of Remote Participants per Continent}
\label{figures:fig4}
\end{figure}

When data are analyzed by continent it is noted that the vast majority of remote participants are from North America, Asia and Europe (Figure \ref{figures:fig4}). This shows a greater engagement by region. Although Brazil appears in third place in the number of remote participants (Figure \ref{figures:fig3}), this does not have significant weight when we analyze the number of participants per continent, where Asia is very close to the United States in the number of remote participants. 

\begin{figure}[ht]
\centering
\includegraphics[scale=1.0]{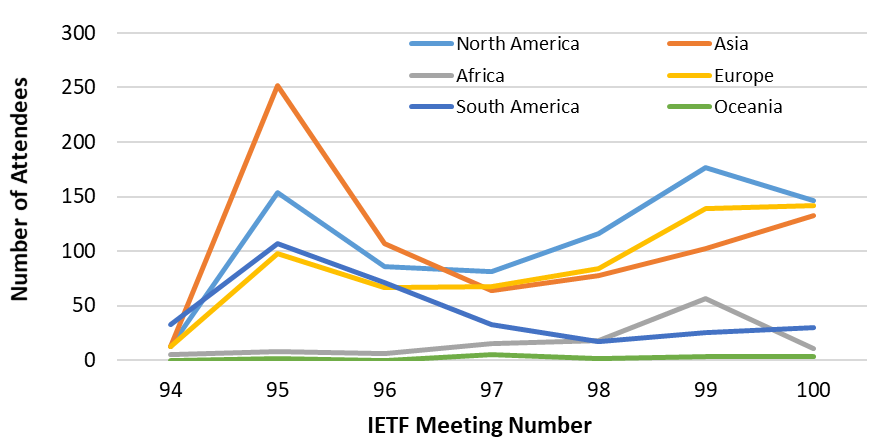}
\caption{Number of Remote Participants per Meeting}
\label{figures:fig5}
\end{figure}

In the history of IETF meetings, there was only a single meeting in South America, the meeting 95 held in Argentina. As the number of remote meetings is still not large (only 8 during the writing of this article), the meeting 95 had a major impact on the overall number of participants from Asia, because at this meeting, probably due to the long trip to South America, we had more than 250 remote Asia participants, 1/3 of all remote Asia participants in a single meeting (Figure \ref{figures:fig4} and Figure \ref{figures:fig5}).

\subsection{South America Participation}

\begin{figure}[ht]
\centering
\includegraphics[scale=1.0]{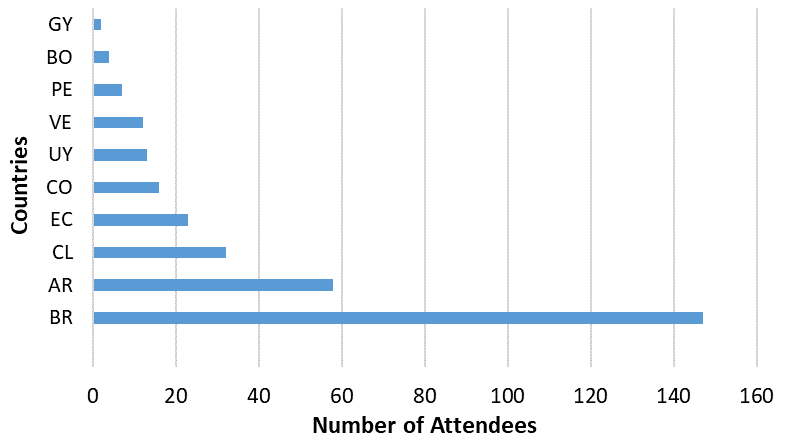}
\caption{South America Remote Participation}
\label{figures:fig6}
\end{figure}

Analyzing the number of remote participants in South America alone, we can note that Brazil has played an important role in relation to the total number of participants. Figure \ref{figures:fig6} shows that Brazil has almost the same number of remote participants as all other South American countries together. It should be noted that Brazil is the most populous country in South America.

\begin{figure}[ht]
\centering
\includegraphics[scale=1.0]{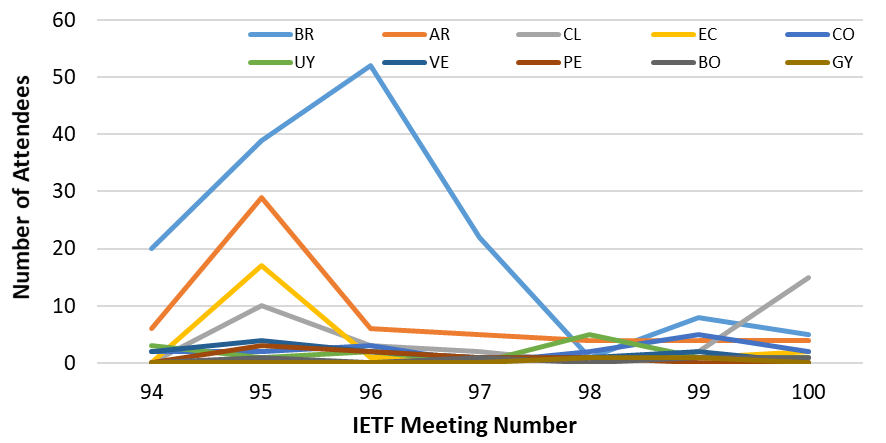}
\caption{South America Remote Participants}
\label{figures:fig7}
\end{figure}

However, after the meeting 96 held in Germany, there is a downward trend in the participation of South American countries. Including Brazil itself (Figure \ref{figures:fig7}). At the meeting 96 we had a peak of remote participants in Brazil. About 60\% of these participants were from the University of Pernambuco (UPE) that organized a Hub to participate remotely in this meeting totaling 22 participants. This behavior shows the impact of remote Hubs. It is important to note that there are some obstacles to remote participation: (1) timezone; (2) appropriate location and a lack of local / regional leadership.

\begin{figure}[ht]
\centering
\includegraphics[scale=1.0]{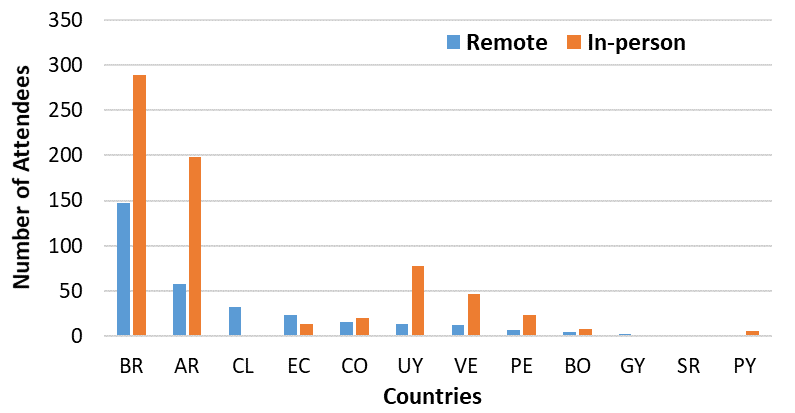}
\caption{Comparison between face-to-face and remote participation in South America}
\label{figures:fig8}
\end{figure}

Analyzing the number of face-to-face participants in South America, we can note that Brazil and Argentina have the largest number of participants (Figure \ref{figures:fig8}), reflecting directly the number of remote participants (Figure \ref{figures:fig7}). Again Brazil stands out in relation to the other countries of South America, having a total of almost 300 participants in the last years. It should be noted that this number of face-to-face presentations refers to meetings with records made available by the IETF (since meeting 72). For more details on participation data in South America, we suggest the efforts made by several authors in \cite{Braga:2014b} and \cite{Braga:2017c}.  

\subsection{Relationship between the number of Drafts / RFCs with the participation of the meetings}

One question regarding remote participation is the degree of participation in IETF documents by remote participants. In other words, do remote participants contribute effectively to the production of Drafts and RFCs?

\begin{figure}[ht]
\centering
\includegraphics[scale=1.0]{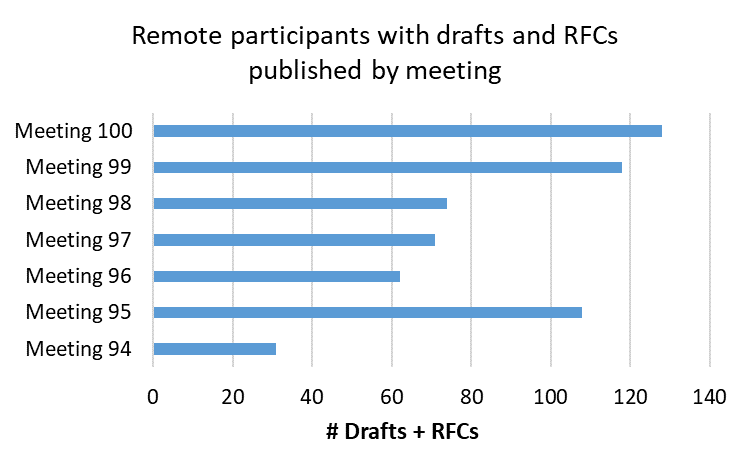}
\caption{Drafts and RFCs published by remote participants}
\label{figures:fig9}
\end{figure}

Figure \ref{figures:fig9} shows that remote participants are increasingly involved in IETF documents. It is possible to see a growing trend of remote participants and their involvement in drafts and RFCs. This demonstrates that the initiative of conveying the meetings is working well and reducing barriers so that the IETF community can collaborate more and more with each other. 

\begin{figure}[ht]
\centering
\includegraphics[scale=.9]{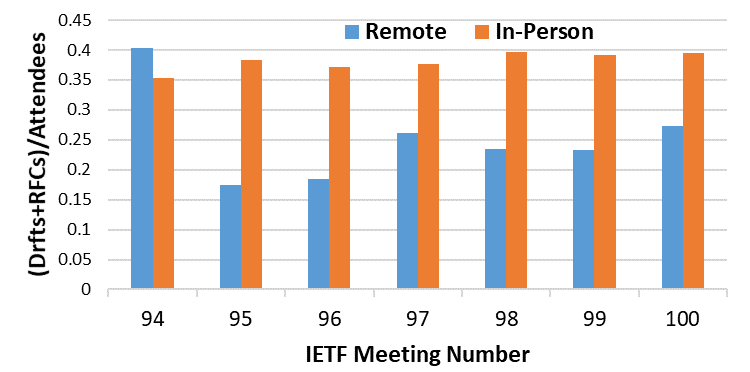}
\caption{Number of drafts and RFCs published by face-to-face participants}
\label{images:fig10}
\end{figure}

\begin{figure}[ht]
\centering
\includegraphics[scale=0.8]{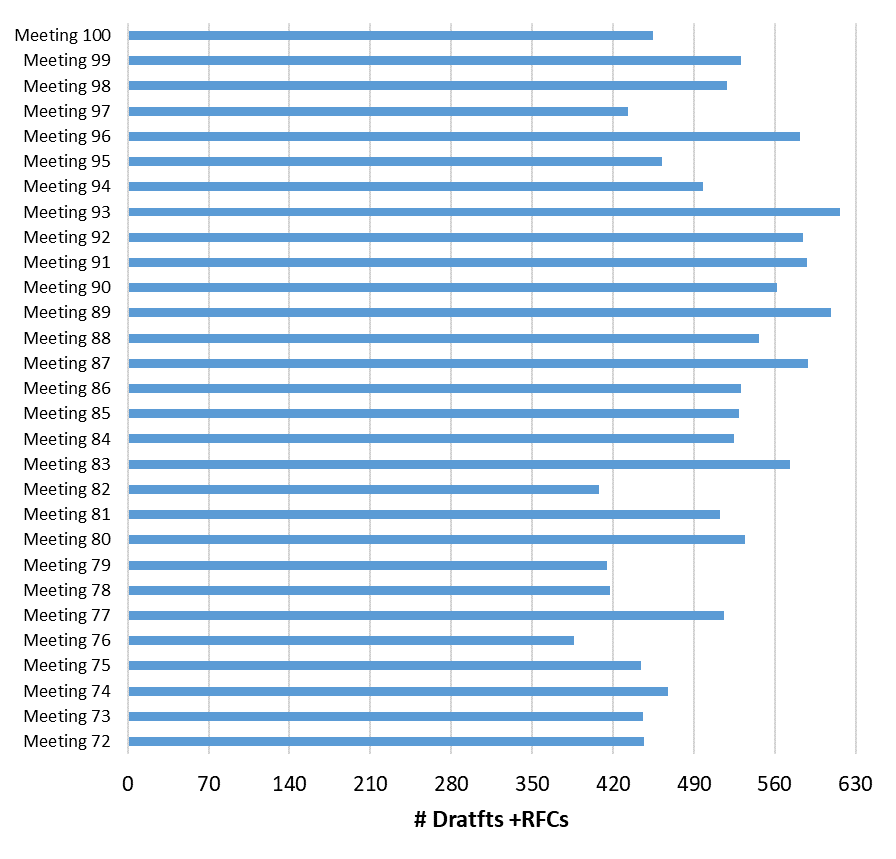}
\caption{Proportion between number of documents and number of participants}
\label{images:fig11}
\end{figure}

Another point to be investigated is the degree of production of face-to-face participants in drafts and RFCs. Figure \ref{images:fig10} shows that there is rather great engagement of members participating in face-to-face meetings in the production of IETF / IRTF documents. Proportionally, the amount of drafts and RFCs produced by remote and face-to-face participants (Figure \ref{images:fig11}), with the exception of meeting 94, shows that IETF face-to-face meeting participants produce an average of twice as much as remote participants. 

\section{Final Considerations}

In this work, data collection was performed on remote and face-to-face participation in IETF meetings through the development of an automated collection tool, from data available on the IETF's pages. A Web Crawler that gathered information from thousands of participants, RFCs, and drafts was implemented in Java.

It was possible to observe that remote participation is growing at each meeting and that this does not necessarily affect the number of attendees. In addition, the data analyzed show that remote participants are increasingly engaged in producing documents within the IETF. Brazil has been losing more and more remote participants, on the other hand, it is still the South American country that has more remote participants. Finally, it was possible to observe that the countries of South America that frequently participate in face-to-face meetings are those that have more remote users at meetings, showing that the number of face-to-face participations is directly related to the number of remote participations. This last point is shown as a positive factor, for example, for programs such as the fellowship offered by the Internet Society.

As future work we intend to analyze the influence of the organizations in the participation of the IETF. Which companies or institutions most encourage members of the IETF community will be one of the points analyzed. Another aspect to be analyzed is whether there are specific groups participating remotely from meetings.

Regardless of results coming from future efforts, we would finally recommend encouraging the more intense participation of the academy by inviting people who can lead local initiatives and maintain a strong interaction among these leaders.

\section{Thanks}

From Juliao Braga: Supported by CAPES – Brazilian Federal Agency for Support and Evaluation of Graduate Education within the Brazil’s Ministry of Education.

\bibliographystyle{sbc}
\bibliography{marcelo}

\end{document}